\documentclass[10pt,conference]{IEEEtran}
\usepackage{cite}
\usepackage{amsmath,amssymb,amsfonts}
\usepackage{algorithmic}
\usepackage{graphicx}
\usepackage{textcomp}
\usepackage{xcolor}
\usepackage{balance}

\usepackage{booktabs}
\usepackage{hyperref}
\usepackage{caption}	 % http://mirror.easyname.at/ctan/macros/latex/contrib/caption/caption-eng.pdf
\newcommand*\rot{\rotatebox{90}}
\graphicspath{ {./} }

\providecommand{\tightlist}{%
  \setlength{\itemsep}{0pt}\setlength{\parskip}{0pt}}

\usepackage{etoolbox}
\usepackage{tcolorbox}
\usepackage{array}

\tcbset{colback=gray!5!white,colframe=gray!85!black}

\AtBeginEnvironment{quote}{\begin{tcolorbox}[leftrule=2mm,bottomrule=0mm,toprule=0mm,rightrule=0mm,boxsep=0mm,grow to right by=-3mm, grow to left by=-3mm]\small}
\AtEndEnvironment{quote}{ \end{tcolorbox}}

\author{
	\IEEEauthorblockN{Wouter Groeneveld\IEEEauthorrefmark{1}, Laurens Luyten\IEEEauthorrefmark{2}, Joost Vennekens\IEEEauthorrefmark{3}, and Kris Aerts\IEEEauthorrefmark{1}} \\
	\IEEEauthorblockA{
		\IEEEauthorrefmark{1}\textit{KU Leuven, Diepenbeek Campus, Department of Computer Science, Belgium}\\
		\IEEEauthorrefmark{2}\textit{KU Leuven, Sint-Lucas Brussels and Ghent Campus, Department of Architecture, Belgium}\\
		\IEEEauthorrefmark{3}\textit{KU Leuven, De Nayer (Sint-Katelijne-Waver) Campus, Department of Computer Science, Belgium}\\
		\{firstname.lastname\}@kuleuven.be
	}
}

\begin{document}

	\title{Exploring the Role of Creativity in Software Engineering}

\maketitle

\begin{abstract}
In order to solve today's complex problems in the world of software
development, technical knowledge is no longer enough. Previous studies
investigating and identifying non-technical skills of software engineers
show that \emph{creative} skills also play an important role in tackling
difficult problems. However, creativity is typically a very vague
concept to which everyone gives their own interpretation. Also, there is
little research that focuses specifically on creativity in the field of
software engineering. To better understand the role of creativity in
this field, we conducted four focus groups, inviting 33 experts from
four nationally and internationally renowned companies in total. This
resulted in 399 minutes of transcripts, further coded into 39 sub-themes
grouped into seven categories: \emph{technical knowledge},
\emph{communication}, \emph{constraints}, \emph{critical thinking},
\emph{curiosity}, \emph{creative state of mind}, and \emph{creative
techniques}. This study identifies the added value of creativity, which
creative techniques are used, how creativity can be recognized, the
reasons for being creative, and what environment is needed to facilitate
creative work. Our ultimate goal is to use these findings to instill and
further encourage the creative urge among undergraduate students in
higher education.
\end{abstract}


\bibliographystyle{IEEEtran}

\hypertarget{introduction}{%
\section{INTRODUCTION}\label{introduction}}

As enterprise software development gets more and more complex,
overcoming big hurdles takes more than just technical knowledge. In a
recent Delphi study, important skills of exceptional software engineers
were identified \cite{groeneveld2020non}. Industry experts agreed that
being creative (e.g., by approaching a problem from different angles) is
an essential problem solving skill, vital to succeed as a software
developer. However, creativity is a multidimensional concept that is not
as easy to define as one might think.

Software engineering (SE) is said to be an outcome of human knowledge
and creativity \cite{bjornson2008knowledge}. Therefore, we wonder:

\begin{itemize}
\tightlist
\item
  \texttt{Q1}: \emph{What exactly is the role of creativity in the world
  of SE?}
\item
  \texttt{Q2}: \emph{What makes one software developer very creative and
  the other less so?}
\end{itemize}

Even though cognitive creativity and creative behavior are
well-researched in the field of psychology
\cite{simonton2000creativity, petkus1996creative}, it is still
difficult to answer these questions specifically for software
development. As Amin et al.~concluded in their systematic literature
review: ``\emph{The research work on creativity in SE is scattered and
scarce}''\cite{amin2017snapshot}. To shed more light on this subject,
we explore the role of creativity using focus groups as a way to gather
qualitative data from experts in the industry. It is important to note
that there are dozens of definitions of `creativity' in literature.
Rather than using an existing definition as a guideline, our intention
is to explore what is understood by the term creativity, according to
software engineers.

Our goal for this research is to contribute to narrowing the gap between
higher education and the requirements of industry, as creativity is
required of engineers to solve complex problems and research has shown
that creativity is currently underrepresented in higher education
computing curricula \cite{groeneveld2020soft}.

The remainder of this paper is divided into the following sections.
Section 2 describes related work, while section \ref{method} clarifies
the utilized methodology. Next, in section 4, we present and discuss the
results of the focus groups, followed by limitations in section 5.
Section 6 concludes this work.

\hypertarget{related-work}{%
\section{RELATED WORK}\label{related-work}}

John Gero describes creative design by comparing it to innovative and
routine design \cite{gero2000computational}. He defines routine design
in computational terms as an activity which occurs when all necessary
knowledge is a priori available: a pre-established procedure can be
followed to come to a design solution. In contrast, he defines
non-routine design by two subgroups: innovative and creative design. In
innovative design the value of the variables directing the procedure to
establish an outcome, are placed outside the intended range. This leads
to design outcomes that are new but still belong to the same class as
their routine progenitors. In creative design one or more new variables
are introduced in the process leading to an all together new class of
design outcomes.

In the field of cognitive psychology, countless creative models have
been proposed in literature, of which Amabile's \cite{amabile1988model}
(\emph{Expertise, Creative Thinking Skills, Motivation}) and Mooney's 4P
\cite{mooney1963conceptual} (\emph{Process, Product, People, Place})
certainly are the most popular. These models have been well-researched
in context of varying fields, including SE \cite{amin2017snapshot}. One
of the disadvantages of starting out with such a model is perhaps the
biased view of creativity, focusing on only one of the four P's instead
of exploring all possibilities, as our intention is.

Requirements engineering is one of the more popular software-related
research areas where creativity has been studied. In an effort to better
understand problem solving in requirements engineering, Cybulski et
al.~discussed creativity with practitioners using focus groups
\cite{cybulski2003understanding}. According to the authors,
`\emph{requirements engineering is well-recognized as a creative problem
solving activity by the systems development community}'. The framework
utilized to guide the focus groups is reminiscent of Csikszentmihalyi's
three-dimensional systems view of creativity
\cite{csikszentmihalyi2014society}: \emph{context} (individual and
social dimensions), \emph{outcome} (the development of creative
information systems), and \emph{process} (supporting the creative
characteristics of problem solving) form the main elements of creative
problem solving. However, requirements engineering is not software
engineering, and therefore, the findings grouped in these three
dimensions might not be readily applicable in context of enterprise
software development.

`\emph{What makes a great software engineer?}' is a question that Paul
Luo Li answered in his 2016 dissertation \cite{li2016makes}. Good
software engineers are essential to the creation of good software.
Instead of asking what makes a software developer creative and the other
less so, Luo Li asked what makes a software developer great and the
other less so. The answer is, among others, the ability to be creative:
applying novel and innovative solutions based on understanding the
context and limitations of existing solutions. In interviews with
industry experts, `\emph{getting a little creative}' is an
often-returning expression. Sadly, the exact role of creativity itself
is never further explored. We can conclude that creativity indeed plays
an important role in solving programming problems, but we still do not
know how creativity is manifested.

In 2019, Anna E. Bobkowska explored creativity techniques in SE using a
specific training-application-feedback cycle
\cite{bobkowska2019exploration}. Since creativity research has produced
more than a hundred different techniques, the question becomes whether
or not any of these techniques are applicable to the field of SE. Seven
techniques were explored: \emph{naive questions} (1; discover hidden
assumptions and implicit knowledge), \emph{reverse brainstorming} (2;
first express criticism, then motivate to improve), \emph{Lunette} (3;
look at the problem at different levels of abstraction), \emph{Chinese
dictionary} (4; a technique to create atypical classifications),
\emph{What if\ldots{}} (5; search for hidden sequences of consequences),
\emph{I could be more creative if\ldots{}} (6; understand personal
obstacles), and \emph{Let's invite him/her} (7; use creativity patterns
of experts in creativity). Participants left the experiment with an
increased appreciation for creativity techniques, claiming that a mix of
these techniques is likely to be useful in practice. However, instead of
starting from certain techniques and matching them to software
development, our aim is to approach creativity with an open mindset and
let the answers come from the participants instead of the literature.

There seem to be few studies that explore creativity in context of the
practice of SE. Instead, we mostly come across theoretical
considerations. Amin et al.~are right: the research work on creativity,
specifically geared towards SE, is indeed scattered and scarce. In order
to approach the research questions with an open mind, we invited 33
experts from the industry to discuss creativity, as explained in the
following section.

\hypertarget{methodology}{%
\section{METHODOLOGY}\label{methodology}}

\label{method}

We conducted four focus groups, collecting information from developers
from different agile software development companies. A focus group is a
small group of experts in a specific field, who brainstorm about a
specific subject in guided open discussions, to gain a better
understanding on the subject at hand \cite{onwuegbuzie2009qualitative}.
In this case, experts are experienced developers and the subject is
creativity.

The focus group method was adapted for SE based on the work of Kontio et
al. \cite{kontio2004using}, as they state it ``\emph{is best suited to
obtain feedback on new concepts, {[}\ldots{]} generating ideas}''. To
simplify the organization and to increase the likely-hood of
participating, each session was held at a separate company. Instead of
using purposive sampling techniques to compose the focus groups, they
were formed by using preexisting groups, namely colleagues. According to
Fern, this is not bias but rather an advantage \cite{fern2001advanced},
since it has the added advantage that participants feel comfortable as
they know each other. Due to the COVID-19 pandemic, the last two
sessions were held online.

\hypertarget{the-focus-group-process}{%
\subsection{The Focus Group Process}\label{the-focus-group-process}}

Expert selection was done by limiting participation to people with a
technical role that come into daily contact with source code, such as
\emph{developers}, \emph{programmers}, \emph{software architects},
\ldots{} To further increase relevance of answers, experts were required
to have at least six years of experience in the field, and should be
intrinsically motivated to partake in the discussion. Participation was
entirely voluntary. Note that, although the job title of many
participating experts is \emph{developer}, this does not mean that their
job is limited to merely implementing pre-set requirements. The four
agile software development companies involved define the role
\emph{developer} as someone who is involved in both analysis, design,
implementation, support, and maintenance phases of a software project.

The conversations were facilitated by the first author, since he has
more than 11 years of experience in the industry. Even if some
researchers claim that familiarity with the topic will introduce bias
\cite{fern2001advanced}, Kontio et al.~state that ``\emph{In the field
of SE, it is very important for the interviewer to have extensive
knowledge of the theme of the interview}'' \cite{kontio2004using}. To
reduce bias as much as possible, each session was audio-recorded,
transcribed, and cross-validated by the second author, whose field of
expertise is architecture rather than SE.

During two hours, open questions are used to elicit opinions,
perceptions, and ideas, which further help us as researchers to
determine the role of creativity in software development. In order not
to influence the results, we chose to refrain from providing a clear
definition of creativity. By not defining the concept ourselves, we try
to find out what the industry might mean when they seek ``creative''
software engineers. The following questions were addressed:

\begin{enumerate}
\def\labelenumi{\arabic{enumi}.}
\tightlist
\item
  \emph{What is the most creative thing you ever did in your job?}
\item
  \emph{How can you see when a colleague is being creative, or not?}
\item
  \emph{What is the most creative project you ever worked on?}
\item
  \emph{How could you measure how creative someone is?}
\item
  \emph{Which creative techniques did you recently employ?}
\item
  \emph{What is the most important reason to be creative?}
\item
  \emph{Which environment do you need to be creative?}
\end{enumerate}

As an ice-breaker, participants first had to complete the sentences
``\emph{as a developer, I am creative when I \ldots{}}'' (1) and
``\emph{as a developer, I am not creative when I \ldots{}}'' (2). After
this individual warm-up exercise, the facilitator gave a brief overview
of Amabile's Componential Model \cite{amabile1988model} (Expertise,
Creative Thinking Skills, Motivation) and Mooney's 4P Model
\cite{mooney1963conceptual} (Process, Product, People, Place) to
increase the awareness of the multi-dimensional aspects of creativity in
general. Next, the above questions were asked in random order, where
participants were grouped in pairs to brainstorm for five minutes,
followed by ten minutes of discussion in group.

\hypertarget{processing-focus-group-data}{%
\subsection{Processing Focus Group
Data}\label{processing-focus-group-data}}

\begin{figure}[h!]
  \captionsetup{belowskip=-5pt}
  \centering
  \includegraphics[width=0.49\textwidth]{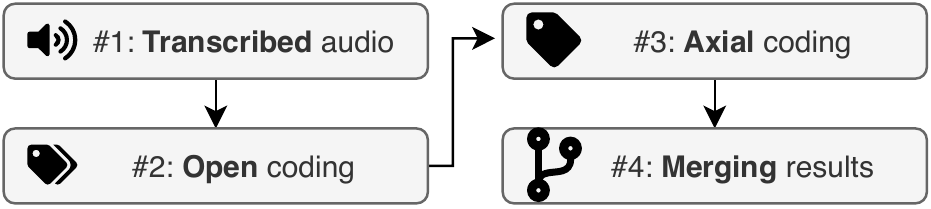}
  \caption{Focus Group Data Processing setps. \label{fig:focusgroupstructure}}
\end{figure}

Data was identified by concepts that `bubble up' and present themselves
after multiple data gathering and reduction steps, as shown in Figure
\ref{fig:focusgroupstructure} and presented in
\cite{onwuegbuzie2009qualitative}. First, focus group recordings were
transcribed (1) by the first author. Then, the transcript and written
notes were read multiple times to apply an open coding step (2),
yielding 82 codes. For this, we followed the guidelines of Richards et
al. \cite{richards2018practical}. All transcripts and notes were
analyzed by the first two authors simultaneously in order to identify
patterns. Afterwards, notes were compared and cross-validated where
needed. Then, codes were further categorized in an axial coding step
(3), yielding 39 subthemes. Multiple revisions of mind maps were made
individually and discussed in group to finally arrive at seven main
themes (4) after merging results from each focus group session.

\begin{figure}[h!]
  \captionsetup{belowskip=-5pt}
  \centering
  \includegraphics[width=0.49\textwidth]{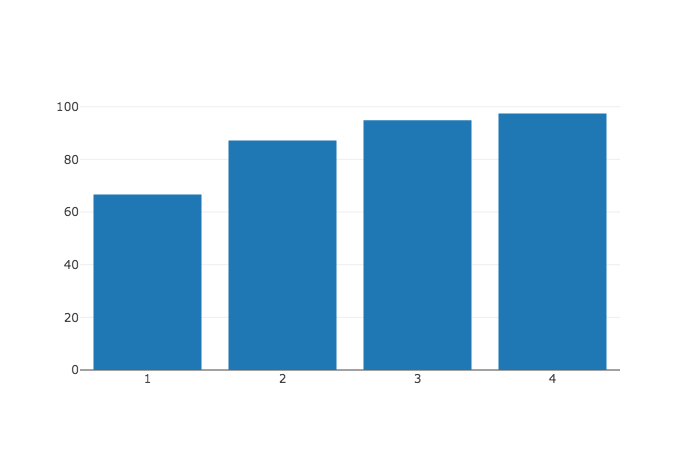}
  \caption{Data saturation: focus groups (x-axis) projected on the percentage of discovered codes (y-axis). Theme distribution rates are also visible in Table \ref{table:composition}. \label{fig:threshold}}
\end{figure}

After conducting the second focus group and merging results, the
calculated thematic data saturation in accordance with
\cite{guest2020simple} was not yet reached (information threshold:
\texttt{27\%}). After the fourth session, the threshold dropped below
the pre-set \texttt{5\%}, indicating the end of the data collection
phase. This effect is visible in Figure \ref{fig:threshold}.

We have used an \emph{emergent-systematic} focus group design
\cite{onwuegbuzie2009qualitative}, in which we organise multiple focus
groups to assess if the themes that emerged from one group also emerge
from other groups. The full list of the 39 identified subthemes can be
consulted in appendix \ref{appendix}.

\hypertarget{results-and-discussion}{%
\section{RESULTS AND DISCUSSION}\label{results-and-discussion}}

\label{discussion}

Table \ref{table:composition} represents focus group meta-data. A total
of 33 participants contributed to 399 minutes of transcripts, yielding a
total of 52.329 words. The average number of participants for each
session was 8, with an average duration of 100 minutes. Companies
involved differed in size, from national to internationally renowned.
This, combined with the participant selection process explained in
section \ref{method}, led us to believe that respondents have
accumulated sufficient experience to be able to contribute to this
research.

\begin{table}[h!]
%\captionsetup{belowskip=-10pt}
  \centering
  \caption{Focus group meta-data, including theme distribution rate to gauge data saturation. \label{table:composition}}
  \begin{tabular}{l r r r r}
    \hline
    Group & Duration & Participants & Words & Theme distr. \\ [0.5ex]
    \hline
    \hline
    Total & 399m & 33 & 52.329 & / \tabularnewline
    \hline
    \#1 & 105m & 7 & 13.454 & 67\% \tabularnewline
    \#2 & 99m & 14 & 12.296 & 87\% \tabularnewline
    \#3 & 100m & 7 & 13.492 & 95\% \tabularnewline
    \#4 & 95m & 5 & 13.087 & 100\% \tabularnewline
  \hline\end{tabular}
\end{table}

\begin{figure*}[h!]
    \centering
    \includegraphics[width=1.0\textwidth]{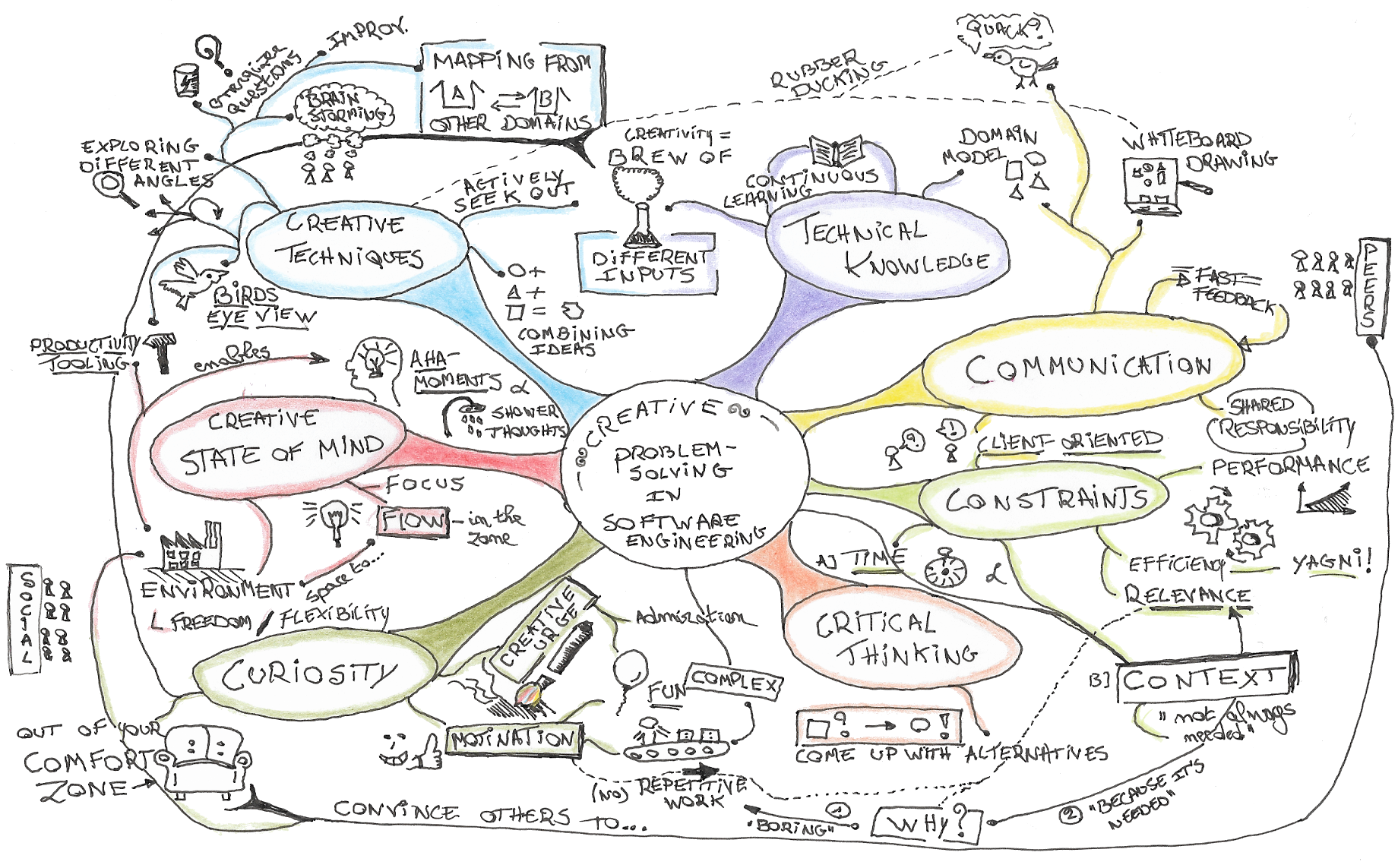}
    \caption{A mind map that summarizes identified themes on creativity. \label{fig:mindmap}}
 \end{figure*}

The seven guidance questions resulted in a lot of similar answers.
Therefore, we opted to group themes in subsets of these questions, each
of which is briefly discussed. The mind map in Figure \ref{fig:mindmap}
was created to provide feedback to respondents on the seven identified
categories and 39 subthemes. It was received very positively and deemed
complete: no new themes were added.

\hypertarget{i-am-not-creative-when-i}{%
\subsection{I am (not) creative when I
\ldots{}}\label{i-am-not-creative-when-i}}

The warm-up exercise yielded surprisingly uniform results, where
developers generally agreed what it means to be creative in their line
of work. A few participants tried to define creativity:

\begin{quote}
`\emph{When someone can come up with an elegant solution to a previously
unsolved non-trivial problem}'
\end{quote}

However, after a brief discussion, it was criticized: it should be
unsolved by the developer, not by anyone, making creativity something
inherently \emph{personal}. Also, it does not need to be elegant: coming
up with an inventive quick but ugly hack that works is also deemed as
creative. The problem at hand should be \emph{complex}; otherwise you
are simply applying previous knowledge to finish a simple task
(cf.~Gero's routine design, \cite{gero2000computational}). Everyone
agreed that they are creative when they model or visualize problems,
split problems into smaller parts, do complex refactoring work, and
brainstorm to approach a problem from another angle.

When are software engineers not so creative? When they blindly take over
code without asking anything, when no alternatives are being devised,
when tightly defined tasks such as simple TODO lists are being ticked
off, when generating boilerplate code, and when \emph{repetition} kicks
in:

\begin{quote}
`\emph{If I just try it until it works, such as fixing dependencies}'
\end{quote}

Engaging in \emph{symptom relief} with quick-fixes, rather than
addressing the root cause, has likewise been mentioned. Participants
also talked about the dangers of \emph{tunnel vision}, where one is not
open to alternatives, consciously or not. Some developers claim that
working on legacy code is not creative, while others state that legacy
code comes with more constraints that require more creative work
compared to maintaining modern projects.

Lastly, interrupts are frequently mentioned as distinctively breaking
the creative flow. It has been proven before that `deep work', being in
the flow of work, is of great help in solving problems
\cite{newport2016deep}.

\hypertarget{requirements-to-be-creative}{%
\subsection{Requirements to be
creative}\label{requirements-to-be-creative}}

In the discussion we identified several requirements or conditions that
are necessary for creativity to take place or to improve it. They range
from personal traits, such as creative skills and sufficient technical
knowledge, to requirements related to the physical, cognitive, and
emotional context of work.

\hypertarget{personal-technical-knowledge-and-creative-skills}{%
\subsubsection{Personal: technical knowledge and creative
skills}\label{personal-technical-knowledge-and-creative-skills}}

There seems to be a thin line between a hack (creative) and fiddling
(ignorance). This means that \emph{technical knowledge} is important, as
also denoted by Amabile's Model \cite{amabile1988model}. Participants
indicated that creative work is impossible for them when exploring a new
framework or programming language. A firm grasp of the basics is
required before being able to do any kind of creative work.

Some participants suggested that creativity says something about the
person, not just about the work they do. They wondered whether a person
with a big creative drive would be creative in any field, given a
presence of necessary basic knowledge. Personality correlations with
creativity have indeed been confirmed before
\cite{reuter2005personality}.

\emph{Curiosity} is cited as an indicator of creativity. A participant
suggested that creativity is constantly looking to improve things.
Whether or not a person is a continuous improver may be inferred from
the frequency of new books on the desk. A `hungry' mind has been proven
before to be a core determinant for achievement \cite{von2011hungry}.

\hypertarget{context-physical-cognitive-and-emotional}{%
\subsubsection{Context: physical, cognitive, and
emotional}\label{context-physical-cognitive-and-emotional}}

\emph{A safe environment}, in which developers have the room to express
themselves and where failing often is not punished. The more people are
penalized for trying out something different, the less new propositions
will emerge from them in the future. A lot of participants felt that
they are more creative when they deliberately step \emph{outside of
their comfort zone}. This could mean presenting something for a
non-technical audience, sharing knowledge, or trying to map problems and
solutions outside of their own problem domain to their own. This
requires a sufficiently safe environment.

\emph{A flexible environment} that enables working whenever and wherever
they want. This reduces stress and leaves more room for the actual
(creative) work - as long as no other mind-numbing activities such as
time tracking are required.

\emph{An interactive environment}. Social interaction, supplemented by
drawing material, increases the flow of information, which in turn makes
that creative brew even more powerful. Sparring with peers of the same
level further eases the exchange of info. Everyone confirmed that higher
levels of creativity are reached together than alone - even if
interaction is not directly created to creative processes: smalltalk and
solutions offered as a joke to entertain each other may just as well
lead to more creative work.

\emph{An environment that facilitates focus.} This includes
noise-canceling headphones and places where you can isolate yourself.
Some stated they are not bothered by background noise or music while
others are easily distracted by it, again hinting at the subjectivity of
the circumstances to work creatively. Landscape offices are reported to
be both a blessing and a curse: the flow of information is high, as are
the amount of unwanted interrupts. Meinel et al.~found that designing
creativity-enhancing workspaces is by no means a simple task
\cite{meinel2017designing}.

\emph{Leveraging productivity tools} allows you to focus on the work
itself, letting the tool do everything else such as boilerplating. Tools
should match your own thinking process. In one focus group, we
accidentally started a Vi versus Emacs war - that ended abruptly with
one participant saying `\emph{Vi won}'.

\hypertarget{motivation-to-be-creative}{%
\subsection{Motivation to be creative}\label{motivation-to-be-creative}}

What motivates developers to be creative? We found several drivers we
grouped in three themes: getting rewarded for their work, being
cognitively challenged, and a personal need to be creative.

\hypertarget{reward}{%
\subsubsection{Reward}\label{reward}}

Although not everyone felt this was absolutely necessary, the
customer-oriented aspect also plays a certain role. A solution to a
problem should be \emph{relevant} for end users, and be received
positively. As one participant expressed:

\begin{quote}
`\emph{Doing something without feedback is just too non-committal, like
working in the rarefield.}'
\end{quote}

Customer focus yields more satisfaction, since clients are the ones who
provide the constraints on which you can unleash your creativity.
Participants also think it is important to be given the responsibility
to do major refactoring work. Appreciation for the work from colleagues
and from clients is also highly regarded.

\hypertarget{challenge}{%
\subsubsection{Challenge}\label{challenge}}

Some developers actively seek out the borders of stepping outside their
comfort zone, while others are forced to get creative by having to
answer difficult ``energizer questions'': in both cases they are
challenged which leads to creative work.

We noticed a similar importance of being challenged when one participant
talked about his creative experience while working on a `brownfield'
legacy conversion project, as opposed to a `greenfield' project where
everything is written from scratch. Everyone agrees that greenfield
projects enable creative thinking, since there is room to do things the
way you want. However, brownfield projects introduce a lot of
\emph{constraints}, such as a strict budget and a legacy database that
needs to be kept online, which in turn triggers more creative work (even
though some developers dislike brownfield projects because they are more
motivated to work on something brand new). Working with constraints has
been mentioned several times during all sessions. Highly constrained
conditions generate more ideas and are generally perceived as more
inspiring \cite{biskjaer2020task}.

Similar challenging aspects can be found in \emph{Full-stack} projects.
They are frequently mentioned as the most creative, involving the
developer \emph{in all of aspects of the design}: from back-end database
structure to front-end UX design. Seeing the project work as a whole,
with all pieces of the puzzle falling into place, seeking out input, and
getting out of your comfort zone. These challenging projects usually
come equipped with lots of constraints, for example a migration or a
hackathon with a limited time frame.

\hypertarget{personal-need}{%
\subsubsection{Personal need}\label{personal-need}}

\emph{Because otherwise it gets boring}. Many developers love their job
precisely because it requires a lot of daily creative work. Assembly
line work was frequently mentioned as being horribly dull. Nonetheless,
someone suggested that a few days of repetitive work might even
stimulate creativity. After mastering something, they start wondering
how to automate things just to give in to their creative urge:

\begin{quote}
`\emph{In a sense, creativity can work therapeutically. I wouldn't be
here if this work wasn't creative. }'
\end{quote}

\hypertarget{tools-and-techniques}{%
\subsection{Tools and techniques}\label{tools-and-techniques}}

To enable and improve creative work, various tools and techniques are
brought forward. They can be categorized under analogies and
(external/internal) feedback.

\hypertarget{analogies}{%
\subsubsection{Analogies}\label{analogies}}

\emph{Mapping solutions from another domain} to your problem field is
one of the more intriguing techniques. For instance, if you are
developing a virtual communication channel, you could try to investigate
how these problems are already solved in the postal system or
face-to-face conversations.

\hypertarget{feedback}{%
\subsubsection{Feedback}\label{feedback}}

As one participant creatively stated:

\begin{quote}
`\emph{Creativity is the brew of different inputs}'
\end{quote}

He actively tried to \emph{seek out these inputs} as much as possible.
Here, external feedback can be obtained directly by asking for it, but
also indirectly by developing your own knowledge relevant for your work.
There are myriad ways to do this, for example:

\begin{itemize}
\tightlist
\item
  Asking for a second opinion from others, not waiting until the code
  review.
\item
  Regularly reading relevant literature to stay up-to-date.
\item
  Attending knowledge sharing meetings.
\item
  Sparring with peers, getting further together.
\end{itemize}

Next to external feedback, it is also possible to develop internal,
self-reflective feedback, such as:

\begin{itemize}
\tightlist
\item
  \emph{Peeling the onion} by keeping on asking `why'.
\item
  Arguing with yourself (\emph{rubber ducking}).
\item
  \emph{Distancing yourself} from your thoughts, thereby approaching the
  problem from other angles.
\item
  Seeking out edge cases like inventing improbable scenarios to
  undermine your own train of thought.
\item
  Switching gears by zooming out to get the broad picture or zooming in
  on one specific aspect of the problem.
\end{itemize}

And of course, brainstorming and modeling the domain on a whiteboard was
mentioned countless times.

\emph{Taking conscious breaks} also came up often. Some developers were
familiar with Hunt's ``Pragmatic Thinking \& Learning''
\cite{hunt2008pragmatic}, in which he characterizes the well-known
`shower thoughts' as asynchronous callbacks from \emph{R}-mode thinking.
Taking a coffee break or deliberately going to the toilet also seemed to
pay off.

\hypertarget{creativity-as-a-value-for-se}{%
\subsection{Creativity as a value for
SE}\label{creativity-as-a-value-for-se}}

Although creativity is considered an important skill in SE, we found
that it is more valued when combined with aspects of work commitment and
critical thinking.

\hypertarget{combined-with-commitment}{%
\subsubsection{Combined with
commitment}\label{combined-with-commitment}}

In one group, there was a heated discussion about bug-fixing. Some claim
finding the bug is creative work, but not fixing it, while others say it
is the other way around. It also seemed to depend on the type of bug,
and who caused it. Of course, it takes some time to understand someone
else's thought process. Some developers do not like to dig deep if they
did not write that piece of code, which is why they dismiss debugging as
not very creative, while others who like to get their hands dirty seem
to claim the opposite. Similar discussions on starting from scratch or
not can be found in `greenfield' versus `brownfield' projects: some
developers dislike brownfield projects because they are more motivated
to work on something brand new. It is clear that the ability to be
creative combined with a commitment to `dig deeper' allows for more
successful work outcomes which employers supposedly value more.

Everyone agreed that being creative is a requirement to successfully
tackle complex problems. Experts noted that most domain-specific
problems are not solved before, as the context and limitations of
problems are almost always unique. Some participants felt the need to
distinguish themselves by committing to be more creative than others.
Others say that creativity happens intuitively:

\begin{quote}
`\emph{Creativity simply arises when you are solving a problem.}'
\end{quote}

Many developers are passionate in what they do. They can unleash their
creative drive at work by showing craftsmanship. However, participants
also mentioned not everyone has this urge, and that is fine too.
Developers can also solve certain problems by simply using their
experience of previously encountered problems, instead of always trying
to be creative. There is a time and place for creativity.

\hypertarget{combined-with-critical-thinking}{%
\subsubsection{Combined with critical
thinking}\label{combined-with-critical-thinking}}

It is important to note that creativity does not always have a positive
connotation. One can come up with extremely creative, but completely
unusable solutions. Participants emphasize the right combination between
creativity and critical thinking, taking into account the context and
constraints of the problem:

\begin{quote}
`\emph{Creativity is the means, not the goal.}'
\end{quote}

\hypertarget{measuring-creativity}{%
\subsection{Measuring creativity}\label{measuring-creativity}}

To jump-start the discussion how to measure creativity, we provided the
example of the interview process when applying for a job in the SE
industry. How can you measure whether an applicant is creative?
According to participants, by \emph{gauging the thought process} when a
problem is presented. We list a selection of the possibilities:

\begin{itemize}
\tightlist
\item
  How are problems approached? Present an impossible scenario and see
  how far they get in trying to solve it.
\item
  Ask questions about something unfamiliar to the applicant.
\item
  Ask open-ended questions outside of the SE field.
\item
  Do a tunnel vision test: ask to list (unit) test cases besides the
  usual suspects. Are all edge cases considered?
\item
  Is the applicant a critical thinker, and if so, does he only utter
  critique or also come up with alternatives?
\item
  The wheel does not always have to be reinvented. Does the applicant
  know the \emph{DRY} and \emph{YAGNI} principles?
\end{itemize}

Participants additionally mentioned puzzling games such as Escape Rooms
and Black Stories as great ways to test creative thinking. Another
option is taking a visual approach by looking at the portfolio of the
applicant. However, the general consensus was that it is difficult to
measure: all indicators are highly subjective.

The same is true for evaluating whether colleagues are creative. Someone
suggested to look at body language. Are they happy, and making a lot of
jokes? Are they `in the zone'? A participant rejected that statement, as
he claimed one can also be very much in the zone by simply `sticking
stamps'. After some discussion, the conclusion was as follows: do they
pause now and then, perhaps thinking? If the pause is too long, they are
stuck. If there is no pause, it is likely to be assembly work and not
creative work. Measuring productivity (visible, doing) is something else
than measuring creativity (invisible, thinking). As one developer
stated:

\begin{quote}
`\emph{Most of my creative work happens in the car. When I'm at work,
all I have to do is type out the solution in my head.}'
\end{quote}

Next to the earlier described curious and `hungry' mind, encouraging
others to think creatively by frequently engaging in the discussion is
also seen as an indicator for creativity. According to participants,
communication is an important aspect of creative work in SE. Everyone
agreed on the following:

\begin{quote}
`\emph{An open-ended question should trigger something in a creative
person.}'
\end{quote}

To conclude, observing certain behavior could be a precondition for
creativity, not a guarantee. It is easier to evaluate whether a product
is creative than whether the process is creative. When asked if
effectiveness is linked to creativity, someone commented that it is not
a requirement, however, in SE, effectiveness is usually an important
constraint.

\hypertarget{limitations}{%
\section{LIMITATIONS}\label{limitations}}

The qualitative nature of this study makes it difficult to provide a
unifying definition of creativity applicable to everyone in SE. However,
defining creativity was never the goal of this research. By adhering to
the methodology and by cross-validating data, as explained in section
\ref{method}, we believe that the results provide relevant and
interesting insights that enlightens the role of creativity.

Also, since creativity is a context sensitive and very personal matter,
generalizations prove to be very challenging. This could be mitigated by
further reproducing this study in other countries and companies. We are
convinced to have collected enough data to contribute to creativity
research.

As mentioned by some participants, only creative people are attracted to
join the focus groups. We intentionally selected on this intrinsic
interest to increase relevance of answers. In this way, we could answer
\texttt{Q1} and \texttt{Q2}. However, it is still unclear how much
creativity is applied daily in the field of SE. Not every discussed form
of creativity (\texttt{Q1}) is seen as desirable or as making one
developer better than the other (\texttt{Q2}). For example, the creative
one who continously looks for new ways without effectively delivering a
final product may be a worse developer.

\hypertarget{conclusions-and-future-work}{%
\section{CONCLUSIONS AND FUTURE
WORK}\label{conclusions-and-future-work}}

The 33 participants of four focus groups brainstormed about the role of
creativity in SE, resulting in seven distinct themes, as displayed in
Figure \ref{fig:focusgroupstructure}: \emph{technical knowledge},
\emph{communication}, \emph{constraints}, \emph{critical thinking},
\emph{curiosity}, \emph{creative state of mind}, and \emph{creative
techniques}. These themes and subthemes also provide insight into how
creativity could be measured, the reasons for being creative, and what
environment is needed to enhance creative work.

There seem to be different levels of creative drive: some software
engineers are more passionate than others, both in their work and
beyond. Regarding the domain-specificity of creativity
\cite{silvia2009creativity}, we feel that most techniques could also be
applied to architecture for example. Of course, both software design and
architecture design share the word \emph{design}. The creative urge
thrives in the unexplored: `\emph{it is the unknown that is the most
creative}'.

We hope this study is useful for two particular groups. First, for SE
practitioners, wo are looking for practical tips on creative problem
solving. As discussed in Section \ref{discussion}, each of the seven
identified creative dimensions can prove to be an effective tool when
facing a problem in the SE world. Second, for the computing education
community, who aspire to inject more creativity into their curriculum.

Our ultimate goal is to imbue and further encourage the creative urge
among undergraduate students in higher education. In future work, we
plan to develop a theoretical framework based on this work and cognitive
psychology literature to assess the creativity of SE students. We aspire
to track and enhance students' progress in creative problem solving
using pre- and post-intervention measurements.

\hypertarget{data-availability}{%
\section{DATA AVAILABILITY}\label{data-availability}}

The transcripts of the focus groups are available upon request. All
identified codes can be inspected in Appendix \ref{appendix}. The
scripts used to analyze the codes and generate the data saturation graph
of Figure \ref{fig:threshold} are available on
\url{https://people.cs.kuleuven.be/~wouter.groeneveld/creafocus/}.

\hypertarget{acknowledgment}{%
\section{ACKNOWLEDGMENT}\label{acknowledgment}}

We wish to thank all participants, who managed to find a few hours time
to creatively brainstorm about creativity, for their valuable
contributions to this research. Thank you all very much!

\appendix

\label{appendix}

Table \ref{fulltable} contains all identified subthemes categorized in
seven themes during the coding process, along with the occurrence in
each focus group. These themes are also visually summarized in the mind
map of Figure \ref{fig:mindmap}.

\begin{table*}
\small
\begin{center}
\begin{tabular*}{\textwidth}{l p{0.65\textwidth}|>{\centering}p{0.05\textwidth}|>{\centering}p{0.05\textwidth}|>{\centering}p{0.05\textwidth}|>{\centering\arraybackslash}p{0.05\textwidth}}
\\
\toprule
 & & \multicolumn{4}{c}{\textbf{Focus Group Occurrence}}
\\
 & \textbf{Subtheme}
 & \textbf{1}
 & \textbf{2}
 & \textbf{3}
 & \textbf{4}
\\
\midrule
  & Employing a \emph{birds eye view} & & \checkmark & &
\\
  & Exploring the problem from \emph{different angles} & \checkmark & & &
\\
  & Asking \emph{energizer questions} to bootstrap the creative process & & \checkmark & &
\\
  & Different forms of \emph{brainstorming} & \checkmark & \checkmark & \checkmark & \checkmark
\\
  & Mapping solutions from \emph{outside the problem domain} & \checkmark & \checkmark & &
\\
  & Actively \emph{seeking out new information} & & & \checkmark &
\\
  \rot{\rlap{\textbf{Crea. Techniques}}}
  & \emph{Combining multiple ideas} into a better one & & \checkmark & \checkmark &
\\
\midrule
  & Creativity as the \emph{'brew' of different inputs} & & & \checkmark &
\\
  & \emph{Continuous learning} & \checkmark & & &
\\
  \rot{\rlap{\textbf{Knowl.}}}
  & Using \emph{domain modeling} to gain new insights & \checkmark & & &
\\
\midrule
  & Using the '\emph{rubber duck}' when stuck & \checkmark & \checkmark & \checkmark &
\\
  & Visualizing the problem on a \emph{whiteboard} & \checkmark & \checkmark & \checkmark & \checkmark
\\
  & Getting \emph{fast feedback} & \checkmark & \checkmark & \checkmark & \checkmark
\\
  & \emph{Sharing responsibility} with all team members &  \checkmark & & &
\\
  & \emph{Client-oriented} designing & & \checkmark & & 
\\
  \rot{\rlap{\textbf{Communicat.}}}
  & Working closely with \emph{peers} &  \checkmark & & & 
\\
\midrule
  & Putting emphasis on \emph{efficiency}  & & & & \checkmark
\\
  & PUtting emphasis on \emph{relevance} & & \checkmark & & \checkmark
\\
  & Putting emphasis on \emph{performance} & & & \checkmark &
\\
  & 'YAGNI': \emph{You Aint Gonna Need It} & \checkmark & & & 
\\
  & \emph{Time-constraints} & \checkmark & \checkmark & \checkmark & \checkmark
\\
  \rot{\rlap{\textbf{Constraints}}}
  & Relevance of \emph{context} & \checkmark & \checkmark & \checkmark & \checkmark
\\
\midrule
  & \textbf{Critical thinking}; coming up with \emph{alternatives} & \checkmark & & \checkmark & 
\\
\midrule
  & Start with '\emph{Why?}' & \checkmark & & & 
\\
  & The dangers and joy of \emph{repetitive work} & \checkmark & \checkmark & \checkmark & \checkmark
\\
  & \emph{Complexity} as a motivational factor & & \checkmark & &
\\
  & Programming '\emph{for fun}'! & \checkmark & \checkmark & &
\\
  & \emph{Admiration} for creative work of others & \checkmark & & &
\\
  & Giving in to that \emph{creative urge} & \checkmark & &  \checkmark &
\\
  & \emph{Motivation} in general & \checkmark & \checkmark & \checkmark & \checkmark
\\
  & \emph{Convincing others} work creatively & & \checkmark & & 
\\
  & Getting out of that \emph{comfort zone} & \checkmark & \checkmark & \checkmark &
\\
  \rot{\rlap{\textbf{Curiosity}}}
  & Creative problem solving as a \emph{necessity} & & \checkmark &  & 
\\
\midrule
  & The \emph{social nature} of software development & \checkmark & \checkmark & & 
\\
  & An environment that facilitates \emph{freedom} & \checkmark & \checkmark & \checkmark & \checkmark
\\
  & An environment that facilitates \emph{flexibility} & \checkmark & & & \checkmark
\\
  & Being in a \emph{creative 'flow'} & \checkmark & \checkmark & \checkmark & \checkmark
\\
  & \emph{'Aha' moments} and shower thoughts & \checkmark & & \checkmark &
\\
  \rot{\rlap{\textbf{State of mind}}}
  & Employing different \emph{productivity tools} & \checkmark & & &
\\
\midrule
\midrule
 & Total percentage of \emph{found} themes in each group & 69.23\% & 56.41\% & 46.15\% & 30.77\%
\\
 & Total percentage of \emph{new} themes in each group & 66.67\% & 20.51\% & 7.69\% & 2.56\%
\\
\bottomrule
\end{tabular*}
\caption{\small The identified themes and subthemes on creativity, with their occurrences in each conducted focus group. The number of emergent themes drop below the pre-set threshold of 5\% after the fourth session, indicating thematic data saturation.   \label{fulltable}}
\end{center}
\end{table*}

	\balance
	\bibliography{report.bib}

\begin{thebibliography}{10}
\providecommand{\url}[1]{#1}
\csname url@samestyle\endcsname
\providecommand{\newblock}{\relax}
\providecommand{\bibinfo}[2]{#2}
\providecommand{\BIBentrySTDinterwordspacing}{\spaceskip=0pt\relax}
\providecommand{\BIBentryALTinterwordstretchfactor}{4}
\providecommand{\BIBentryALTinterwordspacing}{\spaceskip=\fontdimen2\font plus
\BIBentryALTinterwordstretchfactor\fontdimen3\font minus
  \fontdimen4\font\relax}
\providecommand{\BIBforeignlanguage}[2]{{%
\expandafter\ifx\csname l@#1\endcsname\relax
\typeout{** WARNING: IEEEtran.bst: No hyphenation pattern has been}%
\typeout{** loaded for the language `#1'. Using the pattern for}%
\typeout{** the default language instead.}%
\else
\language=\csname l@#1\endcsname
\fi
#2}}
\providecommand{\BIBdecl}{\relax}
\BIBdecl

\bibitem{groeneveld2020non}
W.~Groeneveld, H.~Jacobs, J.~Vennekens, and K.~Aerts, ``Non-cognitive abilities
  of exceptional software engineers: a delphi study,'' in \emph{Proceedings of
  the 51st ACM Technical Symposium on Computer Science Education}, 2020, pp.
  1096--1102.

\bibitem{bjornson2008knowledge}
F.~O. Bj{\o}rnson and T.~Dings{\o}yr, ``Knowledge management in software
  engineering: A systematic review of studied concepts, findings and research
  methods used,'' \emph{Information and Software Technology}, vol.~50, no.~11,
  pp. 1055--1068, 2008.

\bibitem{simonton2000creativity}
D.~K. Simonton, ``Creativity: Cognitive, personal, developmental, and social
  aspects.'' \emph{American psychologist}, vol.~55, no.~1, p. 151, 2000.

\bibitem{petkus1996creative}
E.~Petkus~Jr, ``The creative identity: Creative behavior from the symbolic
  interactionist perspective,'' \emph{The Journal of Creative Behavior},
  vol.~30, no.~3, pp. 188--196, 1996.

\bibitem{amin2017snapshot}
A.~Amin, S.~Basri, M.~F. Hassan, and M.~Rehman, ``A snapshot of 26 years of
  research on creativity in software engineering-a systematic literature
  review,'' in \emph{International Conference on Mobile and Wireless
  Technology}.\hskip 1em plus 0.5em minus 0.4em\relax Springer, 2017, pp.
  430--438.

\bibitem{groeneveld2020soft}
W.~Groeneveld, B.~A. Becker, and J.~Vennekens, ``Soft skills: What do computing
  program syllabi reveal about non-technical expectations of undergraduate
  students?'' in \emph{Proceedings of the 2020 ACM Conference on Innovation and
  Technology in Computer Science Education}, 2020, pp. 287--293.

\bibitem{gero2000computational}
J.~S. Gero, ``Computational models of innovative and creative design
  processes,'' \emph{Technological forecasting and social change}, vol.~64, no.
  2-3, pp. 183--196, 2000.

\bibitem{amabile1988model}
T.~M. Amabile, ``A model of creativity and innovation in organizations,''
  \emph{Research in organizational behavior}, vol.~10, no.~1, pp. 123--167,
  1988.

\bibitem{mooney1963conceptual}
R.~L. Mooney, ``A conceptual model for integrating four approaches to the
  identification of creative talent,'' \emph{Scientific creativity: Its
  recognition and development}, pp. 331--340, 1963.

\bibitem{cybulski2003understanding}
J.~Cybulski, L.~Nguyen, T.~Thanasankit, and S.~Lichtenstein, ``Understanding
  problem solving in requirements engineering: Debating creativity with is
  practitioners,'' in \emph{PACIS 2003: Proceedings of the Seventh Pacific Asia
  Conference on Information Systems}.\hskip 1em plus 0.5em minus 0.4em\relax
  University of South Australia, 2003, pp. 465--482.

\bibitem{csikszentmihalyi2014society}
M.~Csikszentmihalyi, ``Society, culture, and person: A systems view of
  creativity,'' in \emph{The systems model of creativity}.\hskip 1em plus 0.5em
  minus 0.4em\relax Springer, 2014, pp. 47--61.

\bibitem{li2016makes}
P.~L. Li, ``What makes a great software engineer,'' Ph.D. dissertation, 2016.

\bibitem{bobkowska2019exploration}
A.~E. Bobkowska, ``Exploration of creativity techniques in software engineering
  in training-application-feedback cycle,'' in \emph{Workshop on Enterprise and
  Organizational Modeling and Simulation}.\hskip 1em plus 0.5em minus
  0.4em\relax Springer, 2019, pp. 99--118.

\bibitem{onwuegbuzie2009qualitative}
A.~J. Onwuegbuzie, W.~B. Dickinson, N.~L. Leech, and A.~G. Zoran, ``A
  qualitative framework for collecting and analyzing data in focus group
  research,'' \emph{International journal of qualitative methods}, vol.~8,
  no.~3, pp. 1--21, 2009.

\bibitem{kontio2004using}
J.~Kontio, L.~Lehtola, and J.~Bragge, ``Using the focus group method in
  software engineering: obtaining practitioner and user experiences,'' in
  \emph{Proceedings. 2004 International Symposium on Empirical Software
  Engineering, 2004. ISESE'04.}\hskip 1em plus 0.5em minus 0.4em\relax IEEE,
  2004, pp. 271--280.

\bibitem{fern2001advanced}
E.~F. Fern and E.~E. Fern, \emph{Advanced focus group research}.\hskip 1em plus
  0.5em minus 0.4em\relax Sage, 2001.

\bibitem{richards2018practical}
K.~A.~R. Richards and M.~A. Hemphill, ``A practical guide to collaborative
  qualitative data analysis,'' \emph{Journal of Teaching in Physical
  Education}, vol.~37, no.~2, pp. 225--231, 2018.

\bibitem{guest2020simple}
G.~Guest, E.~Namey, and M.~Chen, ``A simple method to assess and report
  thematic saturation in qualitative research,'' \emph{PLoS One}, vol.~15,
  no.~5, p. e0232076, 2020.

\bibitem{newport2016deep}
C.~Newport, \emph{Deep work: Rules for focused success in a distracted
  world}.\hskip 1em plus 0.5em minus 0.4em\relax Hachette UK, 2016.

\bibitem{reuter2005personality}
M.~Reuter, J.~Panksepp, N.~Schnabel, N.~Kellerhoff, P.~Kempel, and J.~Hennig,
  ``Personality and biological markers of creativity,'' \emph{European Journal
  of Personality: Published for the European Association of Personality
  Psychology}, vol.~19, no.~2, pp. 83--95, 2005.

\bibitem{von2011hungry}
S.~Von~Stumm, B.~Hell, and T.~Chamorro-Premuzic, ``The hungry mind:
  Intellectual curiosity is the third pillar of academic performance,''
  \emph{Perspectives on Psychological Science}, vol.~6, no.~6, pp. 574--588,
  2011.

\bibitem{meinel2017designing}
M.~Meinel, L.~Maier, T.~Wagner, and K.-I. Voigt, ``Designing
  creativity-enhancing workspaces: A critical look at empirical evidence,''
  \emph{Journal of technology and innovation management}, vol.~1, no.~1, 2017.

\bibitem{biskjaer2020task}
M.~M. Biskjaer, B.~T. Christensen, M.~Friis-Olivarius, S.~J. Abildgaard,
  C.~Lundqvist, and K.~Halskov, ``How task constraints affect inspiration
  search strategies,'' \emph{International Journal of Technology and Design
  Education}, vol.~30, no.~1, pp. 101--125, 2020.

\bibitem{hunt2008pragmatic}
A.~Hunt, \emph{Pragmatic thinking and learning: Refactor your Wetware}.\hskip
  1em plus 0.5em minus 0.4em\relax Pragmatic bookshelf, 2008.

\bibitem{silvia2009creativity}
P.~J. Silvia, J.~C. Kaufman, and J.~E. Pretz, ``Is creativity domain-specific?
  latent class models of creative accomplishments and creative
  self-descriptions.'' \emph{Psychology of Aesthetics, Creativity, and the
  Arts}, vol.~3, no.~3, p. 139, 2009.

\end{thebibliography}

\end{document}